\documentclass{article}
\usepackage{amsmath}
\usepackage{url}
\usepackage{hyperref}
\usepackage{eufrak}
\usepackage{amsfonts}
\usepackage{amssymb}
\usepackage{amsthm}
\usepackage{color}

\usepackage{tikz}
\usepackage{tikzsymbols}
\usepackage[utf8]{inputenc} 
\usepackage{times}
\usepackage{mathtools}
\usepackage{footnote}
\makesavenoteenv{table}
\usetikzlibrary{arrows,shapes}
\usetikzlibrary{matrix}
\usetikzlibrary{shadows}
\usetikzlibrary{positioning}

\tikzset{
    >=stealth',
    pil/.style={
           ->,
           thick,
           shorten <=2pt,
           shorten >=2pt,}
}

\newcommand{\be}{\begin{equation}}
\newcommand{\ee}{\end{equation}}
\newcommand{\bea}{\begin{eqnarray}}
\newcommand{\eea}{\end{eqnarray}}

\newcommand{\E}{\theta}
\newcommand{\e}{e}

\begin{document}

\title{\bf Lorentz symmetries and primary constraints\\ in covariant teleparallel gravity}

\author{Alexey Golovnev${}^{1}$, Mar\'ia-Jos\'e Guzm\'an${}^{2,3}$\\
{\small ${}^{1}${\it Center for Theoretical Physics, British University in Egypt} }\\
{\small\it{ 11837 El Sherouk City, Cairo Governorate, Egypt} }\\
{\small agolovnev@yandex.ru}\\
{\small ${}^{2}${\it Laboratory of Theoretical Physics, Institute of Physics, University of Tartu,}}\\ 
{\small {\it W. Ostwaldi 1, 50411 Tartu, Estonia}}\\
{\small ${}^{3}${\it Departamento de F\'isica y Astronom\'ia, Facultad de Ciencias, Universidad de La Serena,}}\\
{\small\it Av. Juan Cisternas 1200, 1720236 La Serena, Chile,}\\
{\small mjguzman@ut.ee}
}
\date{}

\maketitle

\begin{abstract}
In this article we explore local Lorentz transformations in theories of gravity based on the teleparallel formalism. For the teleparallel equivalent of general relativity (TEGR), the spin connection plays no role in the equations of motion, and therefore it is possible to simply put it equal to zero with no change in physical quantities, and then the theory is formulated purely in terms of the tetrad field which can be freely chosen in any way. In nonlinear modifications of TEGR, this is a more intricate issue, and vanishing spin connection is then the Weitzenb\"{o}ck gauge choice which imposes restrictions on the choice of tetrad. This has led to considering the so-called covariant formulation of $f(T)$ gravity. We examine the primary constraints arising when passing to the Hamiltonian framework, and compute their algebra. We show that the problems of local Lorentz symmetry breaking still appear in this formulation, even if in a different disguise.
\end{abstract}

\section{Introduction}

Modifications to general relativity (GR) are nowadays a well motivated avenue of research, mainly due to the puzzles of the cosmological dark sector which encompass the dark energy and dark matter riddles. Additionally, the requirement for an early accelerated expansion in the inflationary paradigm, alike the well-known  obstacles in the quest for a quantum theory of gravity, are all additional indications that new physics beyond general relativity must be sought. Among multiple pathways to modify GR, there are additions of scalar, vector or tensor fields, as well as generalisations based on the ordinary curvature tensors of the Levi-Civita connection. Nonetheless, a more radical cut has been surveyed over the last years, which consists of modifications to the spin connection itself through vanishing curvature, but adopting either torsion or non-metricity as the fundamental geometric entities encoding the gravitational dynamics \cite{Jimenez:2019woj}. 

Teleparallelism is almost as old as GR itself. It was proposed by Einstein himself in 1925 by pondering the tetrad (or \textit{vierbein}) as the main dynamical field instead of the metric. His aim was to get a unification of GR and electromagnetism by taking advantage of the six extra components of the tetrad; although later his attempts were proven futile, as they only represent Lorentz transformations that leave the metric invariant \cite{Unzicker:2005in}. Although the teleparallel formalism can be regarded as old, modifications of gravity based on teleparallel geometry have been known for only several decades, starting from one-parameter teleparallel gravity, also known as new general relativity (NGR) \cite{Hayashi:1979qx}. Nonlinear modifications of the teleparallel equivalent of general relativity (TEGR), the most important of which are known as $f(\mathbb T)$ gravity \cite{Ferraro:2006jd,Bengochea:2008gz,Ferraro:2008ey}, have a shorter but very intense story. The model was originally proposed with applications to cosmology in order to explain the early inflationary stage of the universe, however it was later used for explaining its late accelerated expansion. Construction of cosmological solutions in $f(\mathbb T)$ gravity and analysis of cosmological perturbations therein are relatively successful \cite{Hashim:2020sez,Hashim:2021pkq} since the number of dynamical degrees of freedom at the linear level is the same as in GR \cite{Golovnev:2018wbh}. An essential feature appearing at this level is a non-vanishing gravitational slip, which seems to be a generic feature of modified teleparallel models, and might help to test these models with future observations, together with confrontation with astrophysics \cite{Golovnev:2020las}.

Nonetheless, the already mentioned physical predictions have to be taken with great care, as the observed disappearance of dynamical modes casts shadows onto correctness of the linear cosmological analysis. At the nonperturbative level, there is strong evidence supporting at most three extra degrees of freedom. Although the now available Hamiltonian analysis for modified teleparallelism is overly intricate, a few works have dealt with it, and some of them do not fully agree on the number of degrees of freedom (d.o.f.), it is quite clear that the modified theory has more d.o.f. than in GR \cite{Li:2011rn,Ferraro:2018tpu,Ferraro:2018axk,Blagojevic:2020dyq}. This fact alone is already an indication of presence of a strong coupling problem. There is also evidence of dynamical modes at higher order perturbations around Minkowski spacetime, and also with linear perturbations around non-trivial tetrad representation of Minkowski spacetime \cite{Golovnev:2020nln, Golovnev:2020zpv}. 

The appearance of extra d.o.f. in modified teleparallel gravities can be understood from the point of view of loss of local symmetries. Since nonlinear modifications of the TEGR Lagrangian do have equations of motion dependent on the boundary term which is sensitive to local Lorentz transformations of the tetrad field, this symmetry is lost and it is expressed in the new degrees of freedom. In the pure tetrad formulation, which can be viewed as taking the Weitzenböck gauge, the local Lorentz transformations act on the tetrad field only, and those symmetries are generically broken in modifications beyond TEGR. However, a covariant version of modified teleparallel theories has been proposed, in which a Lorentz transformation acts simultaneously on the spin connection and on the tetrad field. However, such transformation does not cure the previously mentioned Lorentz breaking, since the covariant TEGR equations of motion obviously have both types of symmetries: with respect to the simultaneous transformations and with respect to transforming the tetrad (or the spin connection) alone, and the latter symmetry is something what is broken anyway and leads to the non-constant rank of the Poisson brackets' algebra. This fact has generated confusion in the community, creating the common belief that the covariant modified teleparallel gravity cures all the problems associated with breaking the Lorentz invariance, which is however not true.

In this article, we will show that a simultaneous Lorentz transformation of both the tetrad and the spin connection is associated with primary constraints that satisfy a closed algebra corresponding to the algebra of the generators of the Lorentz group. Note that having included this type of transformations in the formalism, it is still possible and perfectly valid to perform Lorentz rotations of only the tetrad field, which can sometimes still be a symmetry of the theory due to the effect of remnant symmetries. However, such pure-tetrad Lorentz transformations are represented by additional primary constraints that satisfy the algebra of the generators of the Lorentz group only in the case that the theory reduces to TEGR. For more general modifications of TEGR, such violation of (pure-tetrad) Lorentz invariance is still present in the covariantised version of the theory, which is reflected in the primary constraints of the pure-tetrad Lorentz transformations not being of the first class type in general.

The structure of this article is as follows. We begin by defining the teleparallel theoretical framework and the primary constraints of the TEGR  theory in Sec.~\ref{sec:theory}. A reminder on the mathematical definition of Poisson brackets and their behaviour in field theory is presented in Sec.~\ref{sec:PB}. The pure-tetrad Lorentz constraints and detailed calculations of their algebra are given in Sec.~\ref{sec:Lorentz}. After this, in Sec.~\ref{sec:covariant} we present the covariant version of teleparallel gravity, the new canonical momenta that induces, and how it modifies the old Lorentz constraints presented in the previous section. The primary constraints appearing from these simultaneous Lorentz transformations of both the tetrad and the spin connection  are exhibited in Sec.~\ref{sec:simult}. Finally, we present discussion and conclusions in Sec.~\ref{sec:concl}.

\section{Pure tetrad TEGR and its primary constraints}
\label{sec:theory}

Traditionally, the connection describing a teleparallel spacetime is explicitly introduced through the tetrad field, even though it is also possible to impose flatness and metricity of a connection by Lagrange multiplier terms in the action \cite{Golovnev:2018red}. For convenience, we consider as the fundamental dynamical variable the co-tetrad field $\E^{a}_{\mu}$ instead, though we will indistinctly call it tetrad, too. The spacetime metric is defined as $$g_{\mu\nu}=\eta_{ab}\E^a_{\mu}\E^b_{\nu}$$ in terms of the Minkowski metric for which our signature convention is $\eta_{ab}=\text{diag}(1, -1, -1, -1)$. However it will not be explicitly used in the main text of our paper, except the fact that the preferred overall sign of the action is determined by the chosen signature. 

The tetrad $\e^{\mu}_{a}$ is defined as the matrix inverse of the co-tetrad by the standard formulae in terms of the matrix components $\theta^a_{\mu}e_a^{\nu}=\delta^{\nu}_{\mu},\quad \theta^a_{\mu}e_b^{\mu}=\delta^{a}_{b}$ which are seldom called a completeness relation. In this paper we will denote tetrad and co-tetrad by different letters, not just distinguishing between them by position of the Latin and the Greek indices. We will need that for making the main expressions more transparent.

In the covariant approach the teleparallel connection is defined as
\be
\Gamma^{\alpha}{}_{\mu\nu} = \e_a^{\alpha}\left(\partial_{\mu}\E^a_{\nu}+\omega^{a}{}_{b\mu} \E^b_{\nu}\right),
\label{flatc}
\ee
where $\omega^{a}{}_{b\mu}$ is a flat spin connection, i.e. a connection that vanishes the Riemann tensor
\be 
\label{Riemann}
R^{a}{}_{b\mu\nu} = \partial_{\mu} \omega^{a}{}_{b\nu} - \partial_{\nu} \omega^{a}{}_{b\mu} + \omega^{a}{}_{c\mu}\omega^{c}{}_{b\nu} - \omega^{a}{}_{c\nu}\omega^{c}{}_{b\mu} = 0.
\ee 
Another important condition is that the nonmetricity tensor also vanishes
\be 
Q_{\mu\alpha\beta} = \partial_{\mu} g_{\alpha\beta} - \Gamma^{\rho}{}_{\mu\alpha} g_{\rho\beta} - \Gamma^{\rho}{}_{\mu\beta} g_{\alpha\rho} = 0,
\ee 
a requirement that, using the definition \eqref{flatc}, can be mapped to the antisymmetry condition $\omega_{[ab]\mu} =0$, so that the differential form $\omega$ takes values in the Lie algebra of the Lorentz group.
The flat and metric-compatible spacetime connection \eqref{flatc} produces a nonzero torsion tensor
\be
T^{\rho}{}_{\mu\nu} = \e^{\rho}_{a}(\partial_{\mu} \E^{a}_{\nu} - \partial_{\nu} \E^{a}_{\mu} + \omega^{a}{}_{b\mu} \E^{b}_{\nu} - \omega^{a}{}_{b \nu} \E^{b}_{\mu}  ),
\label{Ttensor}
\ee 
which is the fundamental geometric quantity of metric teleparallel theories.

In the traditional pure tetrad approach we set the spin connection equal to zero: $$\omega^a_{\hphantom{a}b\mu}=0,$$ which is also known as the Weitzenb\"{o}ck gauge in the covariant language. In this way, the only fundamental variable is the tetrad (or co-tetrad) field. This condition is not a locally Lorentz invariant one, though for the teleparallel equivalent of general relativity (TEGR) the violation comes only in a surface term.\\

The TEGR action $S=\int d^4x\cdot \theta\mathcal{L}$ is given by the following Lagrangian density
\begin{equation}
\label{TEGRact}
\mathcal{L}=\mathbb T=\frac14 T^{\alpha\mu\nu}T_{\alpha\mu\nu}+\frac12 T^{\alpha\mu\nu}T_{\mu\alpha\nu}-T_{\mu}T^{\mu}
\end{equation}
where the torsion tensor and the torsion vector are $T^{\alpha}_{\hphantom{\alpha}\mu\nu}=e^{\alpha}_a(\partial_{\mu}\theta^a_{\nu}-\partial_{\mu}\theta^a_{\nu})\equiv e^{\alpha}_a\partial_{[\mu}\theta^a_{\nu]}$ and $T_{\mu}=T^{\alpha}_{\hphantom{\alpha}\mu\alpha}$ respectively, while  $\theta=\sqrt{-g}$ without indices is the co-tetrad's determinant. The Greek indices are raised and lowered by the spacetime metric, while the Latin ones -- by the Minkowski one. 

The only fundamental variable we are using for now is the tetrad field. Working the quadratic terms in the torsion tensor like $$T^{\alpha\mu\nu}T_{\alpha\mu\nu}=T^a_{\hphantom{\alpha}\mu\nu}T^b_{\hphantom{\alpha}\alpha\beta}\eta_{ab}g^{\mu\alpha}g^{\nu\beta}\quad \mathrm{and}\quad T^{\alpha\mu\nu}T_{\mu\alpha\nu}=T^a_{\hphantom{\alpha}\mu\nu}T^b_{\hphantom{\alpha}\alpha\beta}e^{\alpha}_a e_b^{\mu}g^{\nu\beta},$$ one can explicitly rewrite the Lagrangian density (\ref{TEGRact}) as a quadratic form in velocities:
\begin{equation} 
\label{quadact}
{\mathcal L}=(\partial_{\mu} \E^a_{\nu})(\partial_{\alpha} \E^b_{\beta})\cdot\left(\frac14 e^{\mu}_{[c}e^{\nu}_{d]}e^{\alpha}_{[e}e^{\beta}_{f]} \eta_{ab} \eta^{ce} \eta^{df} +\frac12 e^{\mu}_{[b}e^{\nu}_{c]}e^{\alpha}_{[a}e^{\beta}_{d]}\eta^{cd} - e^{\mu}_{[c}e^{\nu}_{a]}e^{\alpha}_{[d}e^{\beta}_{b]}\eta^{cd}\right),
\end{equation}
where, for convenience, the antisymmetrisation inherent to the torsion tensor is moved from the Greek indices to the Latin ones. This expression (\ref{quadact}) can be compactified even more, obtaining the following alternative formulation of the TEGR Lagrangian \cite{Ferraro:2016wht}
\begin{equation}
L = \dfrac{1}{2}  (\partial_{\mu} \E^{a}_{\nu} ) (\partial_{\rho} \E^{b}_{\lambda} ) \e^{\mu}_{c} \e^{\nu}_{e} \e^{\rho}_{d} \e^{\lambda}_{f} \chi_{ab}{}^{cedf},
\label{Lpremetric}
\end{equation}
where the object
\begin{equation}
\chi_{ab}{}^{cedf} =  \eta_{ab} \eta^{c[d} \eta^{f]e} - 2 \delta_{a}^{[d}\eta^{f][e}\delta^{c]}_{b} - 4 \delta_{a}^{[c} \eta^{e][d}\delta^{f]}_{b}
\end{equation}
is identified as the constitutive tensor for TEGR. 
This form makes it very simple to look for the momenta and the Hamiltonian \footnote{In \cite{Guzman:2020kgh} this approach has been used for the classification of primary constraints in New General Relativity.}. For instance, the canonical momenta can be read as
\be 
\label{piFG}
\Pi^{\mu}_{a} = \dfrac{\partial L}{\partial (\partial_0 \E^{a}_{\mu}) } = \E (\partial_{\rho} \E^{b}_{\lambda} ) e^0_c e^{\mu}_e e^{\rho}_{d} e^{\lambda}_f \chi_{ab}{}^{cedf}.
\ee

It is easily seen from the antisymmetry of the torsion tensor, which is reflected in the antisymmetry of the terms in the Lagrangian (\ref{quadact}) and in the antisymmetry properties of $\chi_{ab}{}^{cedf}$, that the trivial primary constraints
\begin{equation} 
\label{grlikeconst}
\Phi^0_a=\Pi^0_a
\end{equation}
appear, which are analogous to vanishing momenta of the lapse and shift in the ADM action of GR, and therefore produce the diffeomorphism constraints at the level of the secondary ones. Note that the latter are often presented as the primary (and the only ones) constraints in GR, but this is only because the lapse and shift are taken as Lagrange multipliers, though in essence they should be treated as dynamical variables with the primary constraints of vanishing momenta.

It is convenient to rewrite the momenta \eqref{piFG} combining it with a tetrad component such that the resulting object $\Pi^{\mu}_{a} \E^{e}_{\mu}$ has only internal indices. If we perform a temporal and spatial split of the spacetime indices, we obtain the following expression
\be 
\Pi^{\mu}_{a} \E^{e}_{\mu} = \E e^{\lambda}_{f} ( \partial_0 \E^{b}_{\lambda} ) C_{ab}{}^{ef}  + \E ( \partial_i \E^{b}_{\lambda} ) e^{0}_{c} e^{i}_{d} e^{\lambda}_{f} \chi_{ab}{}^{cedf}.
\label{LorentzIntermediate}
\ee 
where $C_{ab}{}^{ef} = e^{0}_{c} e^{0}_{d} \chi_{ab}{}^{cedf}$. Now we see that the velocities appear multiplied by the tensor $C_{ab}{}^{ef}$ which acts as the Hessian, and therefore if we find elements $v^e_a$ belonging to the kernel of the Hessian and multiply the expression \eqref{LorentzIntermediate} by them, we have $v^{a}_{e} C_{ab}{}^{ef} = 0$ and get primary constraints. In this way, one set of the elements of the kernel can be taken as $v_{|g|e}{}^{a}=e^{0}_{e}\delta^{a}_{g}$, with the index $_{|g|}$ used as labelling for the elements. Applying it to the Hessian gives straightforwardly $v_{|g|e}{}^{a}C_{ab}{}^{ef}=0$, and therefore the trivial primary constraints \eqref{grlikeconst} are found.

In order to find the Lorentz constraints, we consider the elements of the kernel whose coefficients are $v_{|gh|e}{}^{a} = 2\delta_{[g}^{a}\eta_{h]e}$ (where the indices ${}_{|gh|}$ are labelling six elements of the kernel), which accomplish 
\be 
v_{|gh|e}{}^{a} C_{ab}{}^{ef} = 2 e^{0}_{c} e^{0}_{d} \eta_{e[h} \chi_{g]b}{}^{cedf} = 0
\ee 
since 
\be 
\eta_{e[c} \chi_{a]b}{}^{gehf} = 2 ( \delta^{h}_{[a}\delta^{f}_{c]} \delta^{g}_{b} + \delta^{g}_{[a}\delta^{h}_{c]} \delta^{f}_{b} + \delta^{f}_{[a}\delta^{g}_{c]} \delta^{h}_{b}  )
\ee 
which is fully antisymmetric in all indices. Multiplying  \eqref{LorentzIntermediate} by $v_{|gh|e}{}^{a}$, gives us the constraints associated with the Lorentz group in the pure tetrad approach
 \cite{Ferraro:2016wht,Blixt:2020ekl}
\begin{equation} 
\label{LorConst}
C_{ab} = \Pi_{[a}^{\mu} \theta^{c}_{\mu}\eta_{cb]} - 2 \theta \left( \partial_i \theta^{c}_{\nu} \right) \left( \e_{[a}^{0} \e_{b]}^{i} \e_{c}^{\nu} + \e_{[a}^{i} \e_{b]}^{\nu} \e_{c}^{0} + \e_{[a}^{\nu} \e_{b]}^{0} \e_{c}^{i} \right).
\end{equation}
This constraint differs from the usual Lorentz constraint of tetrad-based GR $\Pi_{[a}^{\mu}\eta_{b]c} \theta^{c}_{\mu}=0$ \cite{Deser:1976ay,Castellani:1981ue} by a Lorentz non-covariant extra term. Note that our definition of the antisymmetrisation sign is ${\mathcal T}_{[ab]}\equiv{\mathcal T}_{ab}-{\mathcal T}_{ba}$. Note also that it is possible to extend the summation over the index $i$, which most importantly comes through the derivatives $\partial_i \theta^c_{\nu}$, to the time index, too, for the addition to the final result will be just zero. We restrict the summation to only spatial index in order to explicitly show that there is no velocity of the tetrad inside it, and therefore it is indeed a constraint. On the other hand, the summation over $\nu$ can also be restricted to only spatial indices. Finally, note that the relative sign between the two terms depends on whether we take $\mathbb T$ or $-\mathbb T$ as the Lagrangian density (which way is preferable depends on the chosen signature).

This derivation can be made even simpler if we go for the standard representation of the TEGR Lagrangian in terms of the superpotential:
$$\mathcal{L}=\frac12 S_{\alpha\mu\nu}T^{\alpha\mu\nu}=\frac12 S_{\alpha\mu\nu} g^{\mu\rho}g^{\nu\sigma}e^{\alpha}_a(\partial_{\rho} \theta^a_{\sigma}-\partial_{\sigma} \theta^a_{\rho})=S_{\alpha\mu\nu} g^{\mu\rho}g^{\nu\sigma}e^{\alpha}_a\partial_{\rho} \theta^a_{\sigma}$$
which suggests the following expression for the momenta
\begin{equation} 
\label{tetmom}
\Pi^{\mu}_{a}=2\theta S_{\alpha\beta\nu}g^{\beta 0}g^{\mu\nu}e^{\alpha}_a.
\end{equation}
Now, for finding the constraint \eqref{LorConst}, we take the following combination of momenta:
$$\Pi^{\mu}_{[a}\eta_{b]c}\theta^c_{\mu}=2\theta S_{\alpha\beta\nu}g^{\beta 0}e^{\alpha}_{[a} e^{\nu}_{b]}$$
in which only the part of the superpotential
\be  
S_{\alpha\beta\nu}=\frac12 (T_{\alpha\beta\nu}+T_{\nu\beta\alpha}+T_{\beta\alpha\nu})+g_{\alpha\beta}T_{\nu}-g_{\alpha\nu}T_{\beta}
\ee 
that is antisymmetric in the lateral indices $\alpha\leftrightarrow\nu$ plays any role. Therefore we get
$$\Pi^{\mu}_{[a}\eta_{b]c}\theta^c_{\mu}=\theta (T^0{}_{\alpha\nu} + 2\delta^0_{\alpha}T_{\nu})e^{\alpha}_{[a} e^{\nu}_{b]}.$$
Substituting the definitions of the torsion tensor and torsion vector, it gives
\be 
\Pi^{\mu}_{[a}\eta_{b]c}\theta^c_{\mu} = 2 \theta \left( \partial_{\alpha} \theta^{c}_{\nu} \right) \left( \e_{c}^{0} \e_{[a}^{\alpha} \e_{b]}^{\nu} + \e_{[a}^{0} \e_{b]}^{\alpha} \e_{c}^{\nu}  + \e_{c}^{\alpha} \e_{[a}^{\nu} \e_{b]}^{0}  \right),
\ee 
which is unaltered if the summation over $\alpha$ is restricted to spatial indexes only, and therefore it is precisely the constraint (\ref{LorConst}).

One might worry about this approach to finding the momenta, since for reliably concluding from a Lagrangian $L=\frac12 C_{ab} {\dot x}^a {\dot x}^b$ that $P_a=C_{ab} {\dot x}^b$ one needs that the matrix $C_{ab}$ be symmetric. However, one can check that the action does correspond to a symmetric matrix of the quadratic form. Indeed, we can write the $\frac12 S_{\alpha\beta\nu}T^{\alpha\beta\nu}$ form of the action as
\be
L =\frac12 \left( \dfrac12\left[ g^{\alpha\mu}g^{\beta\rho}g^{\nu\sigma} +  g^{\nu\mu}g^{\beta\rho}g^{\alpha\sigma} +  g^{\beta\mu}g^{\alpha\rho}g^{\nu\sigma} \right]  -  g^{\nu\sigma}g^{\mu\rho}g^{\alpha\beta} -  g^{\beta\rho}g^{\mu\sigma}g^{\alpha\nu} \right) T_{\alpha\beta\nu} T_{\mu\rho\sigma}
\ee 
and see that it is indeed symmetric with respect to the exchange  $\alpha\beta\nu\longleftrightarrow\mu\rho\sigma$, and therefore it represents an explicitly symmetric matrix of a quadratic form in terms of the torsion tensor components.
It explains why the presentation of momenta in terms of the superpotential is correct. And people who have some experience with the TEGR or $f(\mathbb T)$ equations of motion, should remember that the superpotential indeed comes out naturally from variation of the action with respect to the torsion tensor components.

Note that the first term in the Lorentz constraint (\ref{LorConst}) is the generator of local Lorentz rotations of a tetrad, as follows from taking the momentum $\Pi^{\mu}_a$ as a generator of variation of $\theta^a_{\mu}$. In case of the tetrad-based GR, the constraint would require just that $\Pi_{[a}^{\mu}\eta_{b]c} \theta^{c}_{\mu}=0$ (a weak equivalence, in the Dirac formalism sense) which satisfies an almost obvious Lorentz algebra. However, it is not the case in TEGR since the Lagrangian density is not locally Lorentz invariant. This is why we have the second term in the constraint. It comes from just a surface term and is not locally Lorentz covariant. Its meaning is that, when acting on an expression with momenta, it must provide the non-invariant part of the transformation of momenta, according to their non-invariant definition. Calculation of algebra of these constraints becomes a bit less elementary; we will come to it later.

Note also that the passage from TEGR to $f(\mathbb T)$ theory changes almost nothing at this level. The $f(\mathbb T)$ action can be represented in the Jordan frame as $\phi{\mathbb T}-V(\phi)$, the effect of which is simply in multiplying  the momenta (\ref{tetmom}) by $\phi$, and therefore the non-covariant term in the Lorentz constraint (\ref{LorConst}) gets multiplied by $\phi$, too. One more primary constraint is vanishing of the scalar field momentum. These two facts drastically alter the constraint algebra of $f(\mathbb T)$ and similar models, and are responsible for the breaking of local Lorentz invariance.

\section{Reminder on Poisson brackets in field theory}
\label{sec:PB}

Since we would like to understand the details of the constraint algebra among the Lorentz constraints previously found, in this Section we will expose in detail the calculation of Poisson brackets in field theory. The calculation in teleparallel theories of gravity has some additional features that are not very frequent in real physical theories, because of the ``pseudo-invariance'' of the TEGR Lagrangian. As some warming up, let us start from a  simple example of Poisson brackets between different quantities in a scalar field theory in $1+1$ dimensions. We simplify the spatial dimension to only one coordinate, so we do not have to deal with spatial indices. Let us consider a scalar field $\phi$ with an associated momentum - $\pi$. We define the fundamental bracket as $\{\pi (x_1),\phi (x_2)\} =-\delta(x_1-x_2)$. As long as no spatial derivatives are involved, all the rest can be found by antisymmetry, linearity over the field of real numbers, and differentiation property which means e.g. that $$\{\pi(x_1), f(\phi (x_2))\}=-f^{\prime}(\phi)\cdot \delta(x_1-x_2)$$ and in particular implies the Leibnitz rule $$\{\pi,g(\phi) f(\phi)\}=g(\phi)\{\pi, f(\phi)\}+\{\pi, g(\phi)\}f(\phi).$$

\subsection{Dealing with spatial derivatives}

To proceed, let us assume we have two functions
\be 
\label{toyfunct}
\mathcal{F}(x_1)\equiv\pi(x_1)\phi(x_1), \ \ \ \ \ \  \mathcal{G}(x_2)\equiv\partial\phi(x_2),
\ee 
where in the latter equality we are using the shorthand notation for $\frac{\partial\phi}{\partial x}(x_2)$. We are interested in the computation of the Poisson bracket $\{\mathcal{F}(x_1), \mathcal{G}(x_2)\}$, which we can naturally approach by taking the appropriate limit of finite differences:
\begin{multline}
\label{findiff}
\{\mathcal{F}(x_1), \mathcal{G}(x_2)\}
 =  \lim_{\Delta x\to 0}\frac{\{\mathcal F(x_1),  \phi(x_2+\Delta x)\}- \{\mathcal F(x_1), \phi(x_2)\}}{\Delta x}\\
 =  -\lim_{\Delta x\to 0}\left(\vphantom{\int}\phi(x_1)\cdot\delta(x_2+\Delta x - x_1)- \phi(x_1)\cdot\delta(x_2 - x_1) \right).
\end{multline}
To see what (in some natural strong operator topology) this operator (\ref{findiff}) is, we act on a smooth test function $\mathcal C(x_1, x_2)$, and denoting this action by the usual abuse of integral notation, we get for the result integrated over the remaining one variable
\begin{multline}
\label{actresult}
\int dx_1 dx_2 \cdot\{\mathcal{F}(x_1), \mathcal{G}(x_2)\} \cdot \mathcal{C}(x_1,x_2)   = -\lim_{\Delta x\to 0} \int dx_2 \frac{\vphantom{\int}\phi(x_2+\Delta x)\cdot  \mathcal{C}(x_2+\Delta x, x_2) - \phi(x_2) \cdot \mathcal{C}(x_2, x_2)}{\Delta x}\\
 =  -\int dx_2 \left(\mathcal{C}(x_2,x_2)\partial\phi(x_2)+\phi(x_2)\left.\frac{\partial {\mathcal C}(x_1,x_2)}{\partial x_1}\right|_{x_1=x_2}\right).
\end{multline}

The relation (\ref{actresult}) defines the generalised function $\{\mathcal{F}(x_1), \mathcal{G}(x_2)\}$ with our quantities (\ref{toyfunct}), which we can formally write as
\begin{equation}
\label{singscalP}
-\{\pi\phi, \partial\phi\}=\partial\phi\cdot\delta (x_1-x_2) + \phi\cdot\mathfrak{D}^{(x_1)}\delta (x_1 - x_2) 
\end{equation}
where by $\delta (x)$ we denote the operator acting from the space of functions to the set of  numbers (or, more precisely in our case, from the space of functions of two variables to the space of functions of one variable) as evaluation 
$$\delta(x-x_0):\quad f \mapsto f(x_0),$$ 
that can be presented as a formal integration of the function $f$ with a  ``$\delta$-distribution'', while $\mathfrak{D}\delta$ is the derivative of the previous operator $\delta$, whose action on $f$ is:  
\be 
\label{Dd}
\mathfrak{D}\delta (x-x_0):\quad f \mapsto \partial f (x_0).
\ee  
The definition \eqref{Dd} means that, considering the action of $\mathfrak{D}^{(x_1)}\delta (x_1 - x_2)$ onto a smooth test function $\mathcal C$ gives as a result $\frac{\partial}{\partial x_1}\mathcal{C}$, evaluated at the locus of $x_1=x_2$. This in turn can be presented as a formal integration of $\mathcal C$ with the ``distribution'' $-\partial^{(x_1)}\delta(x_1-x_2)$, and then the result is multiplied in (\ref{singscalP}) by $\phi$ without it getting differentiated in the last term. Intuitively, it all can be viewed as a natural continuity property of the transformation generated by $\pi\phi$: acting on a derivative $\partial\phi$ it appears to be a derivative of its action on $\phi$ if we look at the Poisson bracket as an operator on the space of test functions.
On the other hand, both terms in the result (\ref{actresult}) of acting by the operator (\ref{singscalP}) onto a test function $\mathcal C$ can be written together as $-\partial\delta$ integrated with $\phi\mathcal C$.

Let us also expose this computation from the viewpoint of calculating the brackets of smeared quantities, which should have a more regular behaviour. We define the smeared versions of the $\mathcal{F}(x_1)$ and $\mathcal{G}(x_1)$ functions (\ref{toyfunct}) as 
$$\mathfrak{F}(x_1)=\int dz_1 C_{x_1}(z_1)\mathcal{F}(z_1)=\int dz_1 C_{x_1}(z_1) \pi(z_1)\phi(z_1)$$ and 
$$\mathfrak{G}(x_2)=\int dz_2 C_{x_2}(z_2)\mathcal{G}(z_2)=\int dz_2 C_{x_2}(z_2) \partial\phi(z_2),$$
respectively, with arbitrary smearing functions $C_{x_i}(z_i)$ which are only assumed to be smooth enough and localised (have their supports) in the neighbourhoods of the $x_i$ points. One possible choice would be $C_x(z)=\frac{1}{\sqrt{\pi\alpha^2}}e^{-\frac{(z-x)^2}{\alpha^2}}$, which in the limit of $\alpha\to 0$ (in dimension one) brings the quantity back to the unsmeared one and to the case of the bracket (\ref{singscalP}), that has a singular behaviour.  

With the standard definition of Poisson brackets for field theory
$$\{\mathfrak{F}(x_1), \mathfrak{G}(x_2)\}\equiv-\int dy\left(\frac{\delta\mathfrak{F}}{\delta\pi}(y)\cdot \frac{\delta\mathfrak{G}}{\delta\phi}(y)-\frac{\delta\mathfrak{F}}{\delta\phi}(y)\cdot \frac{\delta\mathfrak{G}}{\delta\phi}(y)\right),$$
for the Poisson brackets of the smeared quantities (\ref{toyfunct}) we obtain
\begin{equation} 
\label{smooscalP}
\{\mathfrak{F}(x_1), \mathfrak{G}(x_2)\}=\int dy \cdot C_{x_1}(y)\phi(y) \cdot \partial C_{x_2}(y)=-\int dy \cdot C_{x_2}(y) \cdot \partial\left(\vphantom{\int}C_{x_1}(y)\phi(y)\right)
\end{equation}
which is the correct result for the Poisson bracket of the smeared quantities in consideration. We performed one integration by parts in order to find $\frac{\delta\mathfrak{G}}{\delta\phi}$, and consequently another integration by parts is performed in the obtained expression for the Poisson bracket \eqref{smooscalP} to be easily comparable with the previous result  \eqref{singscalP}. It is not just accidental because, in a sense, the second integration by parts is a compensation for the integration by parts which we were forced to perform for calculating the variational derivative of $\mathfrak G$ (\ref{actresult}). 

Note that the new result \eqref{smooscalP} is fully in accordance with our calculation in terms of the derivative of the $\delta$-functional. We just have to take the $\delta$-and-$\mathfrak{D}\delta$ operator (\ref{singscalP}), act with it upon the test function $C_{x_1}(z_1)C_{x_2}(z_2)$, and integrate the result. It reduces to integration over just one variable (strictly speaking, the initial integration sign of $\int dz_1 dz_2$ stays there again in the usual impressionistic sense when dealing with $\delta$-like objects, meaning integration over the only one variable which remains active after equating the two variables which were there before having been acted by the operator):
\begin{multline*}
\{\mathfrak{F}(x_1), \mathfrak{G}(x_2)\}=\int dz_1 dz_2\cdot  \{\mathcal{F}(z_1), \mathcal{G}(z_2)\}\cdot C_{x_1}(z_1)C_{x_2}(z_2)\\
=-\int dy \left((\partial\phi(y))C_{x_1}(y)C_{(x_2)}(y)+\phi(y)C_{x_2}(y)\partial C_{x_1}(y)\right)=-\int dy \cdot C_{x_2}(y)  \partial\left(\vphantom{\int}C_{x_1}(y)\phi(y)\right)
\end{multline*}
which indeed coincides with the result (\ref{smooscalP}) of calculating the brackets of the smeared quantities.

What we have seen up to now is that if one quantity in a Poisson bracket is a spatial derivative of some variable, then the variation of another quantity in the same bracket gets differentiated (in the finite differences case, or with the two natural integrations by parts for the calculation with smeared quantities). Note also that it is not the case for another factor in the quantity with the derivative. Indeed, for example, in the case of $\tilde{\mathcal G}=f(\phi)\cdot \partial \phi$ the new factor of $f$ will directly go evaluated at the point $x_2$ when finding the limit of finite differences, and therefore the function $f$ will not produce new $\partial\phi$ derivatives in an analogue of the bracket (\ref{singscalP}):
$$-\{\pi\phi, f(\phi)\partial\phi\}=\left(f^{\prime}(\phi)\phi+f(\phi)\right)\partial\phi\cdot\delta (x_1-x_2) + f(\phi)\phi\cdot\mathfrak{D}^{(x_1)}\delta (x_1 - x_2) $$
This is good news since it means that in such calculations we can use the previous results by applying the Leibnitz rule instead of repeating everything from the very beginning.  In particular, for the bracket with a product of two quantities, $f(\phi)$ and $\partial\phi$, we've got $$\{\mathcal{F} , f(\phi)\cdot \partial \phi\}=\{\mathcal{F} , f(\phi)\}\cdot \partial\phi+f(\phi)\cdot\{\mathcal{F} , \partial \phi\}.$$ 
Analogously to the finite differences case, the smeared Poisson bracket upon integration by parts will not have any new derivative of $f$, on top of what is there due to just the simple $\{\pi,f(\phi)\}$ bracket, since the variation of the smeared $ \tilde{\mathfrak{G}}$
will be given by $-\partial\left(f(\phi(y))C_{x_2}(y)\right)$ which loses the derivative of $f$ upon our subsequent integration by parts, plus an obvious term with the direct variation of the argument of $f$.

\subsection{Effects of boundary terms}

One more issue related to teleparallel gravity which we would like to discuss further through simple examples is the influence of boundary terms. In Lagrangian mechanics, we can play with boundary terms for free, as long as we do not care about global issues, however the Hamiltonian mechanics gets some modifications in its appearance when including total time derivatives in the boundary terms.

If we add just a spatial surface term, like for example $\partial_i \partial^i {\dot \phi}^n$, the definition of momentum does not get modified even when the extra term contains a velocity inside, let alone the value of the Hamiltonian obtained by integrating over the whole space. More accurately, this statement fully holds only if not considering global issues related with physics at the boundary, and assuming that integration by parts is always allowed to be performed. However, the  addition of a full time derivative to the Lagrangian density does make a change in our Hamiltonian description of a given model because it changes the definition of momenta.

The notion of symmetries is more subtle in the Hamiltonian formulation. For example, restricting ourselves for simplicity to pure mechanics, if we take a Lagrangian of two fields as
$$L(\phi,\psi)=\frac12 {\dot\phi}^2,$$
it is then a trivial example of symmetry with respect to arbitrary changes of the ``pure gauge'' variable $\psi$. However, in the Hamiltonian formalism we get the constraint of $P_{\psi}=0$ making this sector not fully free of anything, even though $\psi$ itself is still fully arbitrary while $\phi$ gets a linear dependence on time. The Hamiltonian can be taken to be
$$H=\frac12 P^2_{\phi}+\lambda P_{\psi}.$$
However, since the velocities can not be uniquely solved for, this form of the Hamiltonian is not unique. We can add any function of $P_{\psi}$ as it does not change anything. This is not to say that it can never be a problem. Even for a non-degenerate Lagrangian, if it is of a higher order in velocities than the usual quadratic one, the solution for the velocities in terms of momenta might often be not unique, therefore producing a set of really different Hamiltonians. 

Let us now spoil the symmetry at the level of the Lagrangian without changing the equations of motion:
\be 
\label{toym1}
L(\phi,\psi)=\frac12 {\dot\phi}^2+{\dot\phi}\psi+{\dot\psi}\phi.
\ee
It yields a new Hamiltonian system with
\be
H=\frac12 (P_{\phi}-\psi)^2+\lambda (P_{\psi}-\phi)
\ee
and equations of motion
\bea 
\dot{\psi}=\lambda, & \ \ \  &\dot{\phi} = P_{\phi } - \psi \\
\dot{P}_{\phi} = \lambda, & \ \ \ & \dot{P}_{\psi} = P_{\phi} - \psi.
\eea 
The constraint is unique and therefore first class, with no restriction on the Lagrange multiplier. Therefore, $\psi$ has  anyway full freedom in the time coordinate, however since $\dot{\psi}-\dot{P}_{\phi}=0$, we see that $P_{\phi}-\psi$ must be constant, which can be interpreted as some non-trivial connection between the physical $\phi$ and the pure gauge $\psi$ sectors. After this, we obtain that $\phi$ as a linear function of time, as it should be. In other words, even though we had not destroyed the physical equivalence of the Lagrangian and the Hamiltonian formalisms, the latter became more involved in its mathematical formulation.

Another issue of ``pseudo-invariant'' Lagrangians being treated in the Hamiltonian language, is that even the gauge invariance can be realised in a very non-trivial way. If the definition of momenta lacks covariance, the shape of constraints must take care of that describing also the non-covariant part of the transformation of momenta. Let us illustrate it again at the level of first class constraints in classical mechanics, but with a bit less trivial example.

Let us consider a different Lagrangian
\be 
\label{toym2}
L(\phi,\psi)=\frac12( {\dot\phi}_1-{\dot\phi}_3)^2+\frac12( {\dot\phi}_2-{\dot\phi}_4)^2\ee
which has two ``gauge'' invariances for sums of two $\phi$-s, in the same way as the invariance under changes of $\psi$ \eqref{toym1}. Finding the Hamiltonian is quite trivial a task. Let us just state the system's two first class constraints $P_1+P_3=0$ and $P_2+P_4=0$ which are clearly first class.

Now, we modify the Lagrangian, again without changing equations of motion:
$$L(\phi,\psi)=\frac12( {\dot\phi}_1-{\dot\phi}_3)^2+\frac12( {\dot\phi}_2-{\dot\phi}_4)^2+\frac{d}{dt}f(\phi_1,\phi_2,\phi_3,\phi_4).$$
Each of the new momenta $P_i$ gets added the term  $f_i\equiv\frac{\partial f}{\partial\phi_i}$. The new constraints which follow from the modified Lagrangian, $P_1+P_3-f_1-f_3=0$ and $P_2+P_4-f_2-f_4=0$, do not respect the invariance of the action if the function $f$ does not, because of the non-covariance of the newly defined momenta. However, it is very easy to see that they still have zero Poisson bracket with each other, therefore preserving their first class character, due to the symmetry of the second partial derivatives of $f$, and do not produce any secondary constraints.

\section{The oddities of the Lorentz algebra}
\label{sec:Lorentz}

Now we come back to the issues of teleparallel gravity. The Lagrangian formulation is not invariant under the local Lorentz rotations of the tetrad. Equations of motion are covariant still, because the non-invariance is purely in a surface term. However, it changes the definition of momenta, and we have to deal with effects analogous to the ones we discussed in the case of simple toy models.

\subsection{The basic brackets}

To discuss the issues of Lorentz algebra, let us define the quantities which are parts of the Lorentz constraint (\ref{LorConst}): $\mathcal{C}_{ab}=\mathcal{F}_{ab}-\mathcal{G}_{ab}$ with 
$$\mathcal{F}_{ab}=\Pi_{[a}^{\mu}\eta_{b]c} \theta^{c}_{\mu}=\Pi_{a}^{\mu} \theta^{c}_{\mu}\eta_{cb}-\Pi_{b}^{\mu} \theta^{c}_{\mu}\eta_{ca},$$
$$\mathcal{G}_{ab}= 2 \E \left( \partial_i \theta^{c}_{\nu} \right) \left( \e_{[a}^{0} \e_{b]}^{i} \e_{c}^{\nu} + \e_{[a}^{i} \e_{b]}^{\nu} \e_{c}^{0} + \e_{[a}^{\nu} \e_{b]}^{0} \e_{c}^{i} \right)$$
and study their Poisson brackets. This goes the same way as our toy examples with a scalar field, with the same issue of having spatial derivatives inside the part of $\{\mathcal{F},\mathcal{G}\}$.

The story of $\mathcal{F}_{ab}$ alone is the same as would have been there for just the tetrad formulation of GR with no problems of ``pseudo-invariance'', and is quite elementary. With the usual definition of the fundamental Poisson bracket $\{\Pi^{\mu}_a, \theta^b_{\nu}\}=-\delta_a^b\delta^{\mu}_{\nu}\delta(x_1-x_2)$, we easily get (omitting the overall $\delta$-function factor)
$$\{{\mathcal C}_{ab}, \theta^{d}_{\mu}\}=\{{\mathcal F}_{ab}, \theta^{d}_{\mu}\}=-\delta^d_{[a}\eta_{b]c} \theta^c_{\mu} = -\delta^d_a \theta^c_{\mu}\eta_{cb}+\delta^d_b \theta^c_{\mu}\eta_{ca},$$ $$\{{\mathcal C}_{ab}, e^{\mu}_{d}\}=\{{\mathcal F}_{ab}, e^{\mu}_{d}\}=e^{\mu}_{[a}\eta_{b]d}=e^{\mu}_{a}\eta_{db}-e^{\mu}_{b}\eta_{da},$$ and also $\{{\mathcal C}_{ab},\theta\}=\{{\mathcal F}_{ab},\theta\}=0$. Those are nothing but the simple Lorentz transformations of tensors with upper or lower Lorentz indices, respectively. Having two indices instead of one (like in $e^{\nu}_d e^{\rho}_e$ or in $e^{\nu}_d \theta_{\rho}^e$) will make four terms instead of two, which will cancel each other in case of contracted indices, of course. Analogously, we immediately get the following (standard Lorentzian) algebra:
\begin{equation}
\label{standLor} 
\{ {\mathcal F}_{ab}, {\mathcal F}_{cd}\}
= \eta_{ad}{\mathcal F}_{bc} + \eta_{ac}{\mathcal F}_{db} + \eta_{bd}{\mathcal F}_{ca} + \eta_{bc}{\mathcal F}_{ad}=-\eta_{\left[\vphantom{\frac{A}{B}}a[c\right. }\mathcal{F}_{\left.b\vphantom{\frac{A}{B}}\right]d]},
\end{equation}
where in the last equality the antisymmetrisations are assumed over the pairs $a\leftrightarrow b$ and $c\leftrightarrow d$.

Up to now, all these are trivial results; the expression ${\mathcal F}_{ab}$ acts as a generator of Lorentz transformations, and what we observe is just the appropriate transformation properties. However, due to the Lagrangian density not being covariant, the definitions of momenta are not covariant either, and we have the extra term in $C_{ab}$ which does not transform covariantly, due to the $\partial_i\theta^c_{\nu}$ object.

To see what happens with the algebra when considering this term, we need to find the Poisson bracket with the spatial derivatives. One can proceed the same way as for the scalar fields and get
\begin{equation} 
\label{dertetP}
\{\Pi^{\mu}_{[a}(x_1)\eta_{b]c} \theta^c_{\mu}(x_1), \partial_i \theta^d_{\nu}(x_2)\}=-\delta^d_{[a} \eta_{b]c}\cdot \left(\vphantom{\int}(\partial_i\theta^c_{\nu})\cdot\delta(x_1-x_2)+\theta^c_{\nu}\cdot\mathfrak{D}_i^{(x_1)}\delta(x_1-x_2)\right)
\end{equation}
for our new building block $\{\mathcal{F},\partial\theta\}$. The same result is obtained by defining the bracket in terms of smeared quantities:
\begin{multline*} 
\left\{\int dz_1C_{x_1}(z_1)\Pi^{\mu}_{[a}(z_1)\eta_{b]c} \theta^c_{\mu}(z_1)\ , \int dz_2 C_{x_2}(z_2)\partial_i \theta^d_{\nu}(z_2)\right\}
= \delta^d_{[a} \eta_{b]c}\int dy \cdot C_{x_1}(y)\theta^c_{\nu}(y)\partial_i C_{x_2}(y)\\
=-\delta^d_{[a} \eta_{b]c}\int dy \cdot  \left(\vphantom{\int}C_{x_1}(y)C_{x_2}(y)\cdot\partial_i \theta^c_{\nu}(y)+(\partial_i C_{x_1}(y)) C_{x_2}(y)\cdot \theta^c_{\nu}(y)\right).
\end{multline*}
The first term in the last line transforms the gradient of the tetrad the same way as the generator $\mathcal{F}_{ab}$ transforms the tetrad itself, this is the global part, and it behaves perfectly in a covariant way. The second term with the $\partial C_{x_1}$ contribution to the smeared bracket, or the $\mathfrak{D}\delta$ term in the singular bracket, corresponds to an extra, non-covariant change of the quantity under a local Lorentz transformation which comes due to differentiation of the Lorentz matrix employed for rotating the tetrad.

\subsection{The algebra}

We are now ready to tackle the computation of the full Lorentz algebra of pure tetrad TEGR. As was mentioned before, the $\{\mathcal{F},\mathcal{F}\}$ part is the same as in tetrad GR, while the $\{\mathcal{G},\mathcal{G}\}$ contribution is obviously zero.
To find the algebra of Lorentz constraints, we need one more Poisson bracket:
\be 
\{\mathcal{F}_{ab} , \mathcal{G}_{mn}\}=\left\{\Pi_{[a}^{\mu}\eta_{b]c} \theta^{c}_{\mu}\ , 2 \theta \left( \partial_i \theta^{d}_{\nu} \right) \left( \e_{[m}^{0} \e_{n]}^{i} \e_{d}^{\nu} + \e_{[m}^{i} \e_{n]}^{\nu} \e_{d}^{0} + \e_{[m}^{\nu} \e_{n]}^{0} \e_{d}^{i} \right)\right\}.
\ee 
The determinant $\theta$ has zero Poisson bracket with $\mathcal{F}_{ab}$, and the remaining computation can be regarded as a linear combination of brackets of the general form 
\be 
\{\mathcal{F}, (\partial\theta)eee\}=\{\mathcal{F}, \partial\theta\}eee+(\partial\theta)\{\mathcal{F}, e\}ee+(\partial\theta)e\{\mathcal{F}, e\}e+(\partial\theta)ee\{\mathcal{F}, e\},
\ee 
for each one of the three antisymmetrised $eee$ products with different combinations of indices. 

If to neglect the ${\mathfrak{D}}\delta$ part of the $\{\mathcal{F}, \partial\theta\}$ bracket (\ref{dertetP}), it obviously reproduces the Lorentz algebra again:
$$\{ {\mathcal F}_{ab}, {\mathcal G}_{mn}\}
= (\eta_{an}{\mathcal G}_{bm} + \eta_{am}{\mathcal G}_{nb} + \eta_{bn}{\mathcal G}_{ma} + \eta_{bm}{\mathcal G}_{an})\cdot\delta + \mathrm{the\ part\ of\ }{\mathfrak{D}}\delta$$
since every quantity in $\mathcal{G}_{mn}$ goes then with the correct transformation (as long as we ignore the ${\mathfrak{D}}\delta$ indeed); the transformation terms related to the repeated index $d$ just cancel each other, while all the others give the usual Lorentz transformation rule for the indices $m$ and $n$.
A problem already here is that, unlike in the  $\{\mathcal{F}_{ab} , \mathcal{F}_{mn}\}$ part, the algebra now comes twice, from $\{\mathcal{F}_{ab} , \mathcal{G}_{mn}\}$ and from $\{\mathcal{G}_{ab} , \mathcal{F}_{mn}\}$:
\begin{equation} 
\label{newtermsP}
\{ {\mathcal F}_{ab}, {\mathcal G}_{mn}\}+\{\mathcal{G}_{ab} , \mathcal{F}_{mn}\}
=2( \eta_{an}{\mathcal G}_{bm} + \eta_{am}{\mathcal G}_{nb} + \eta_{bn}{\mathcal G}_{ma} + \eta_{bm}{\mathcal G}_{an})\cdot\delta + \mathrm{the\ parts\ of\ }{\mathfrak{D}}\delta.
\end{equation}

The extra term of the ${\mathfrak{D}}\delta$ operator is yet another problem which makes the following naively unwanted addition to the bracket under consideration, $\{\mathcal{F}_{ab} , \mathcal{G}_{mn}\}$:
\be 
\label{Ddterm}
-2\theta(e^i_{\left[\vphantom{\frac{A}{B}}a\right.} \eta_{\left.\vphantom{\frac{A}{B}}b\right][m}e^0_{n]}+e^0_{\left[\vphantom{\frac{A}{B}}a\right.} \eta_{\left.\vphantom{\frac{A}{B}}b\right][n}e^i_{m]})\mathfrak{D}_i\delta=2\theta\eta_{\left[\vphantom{\frac{A}{B}}a[m\right.}(e^i_{\left.\vphantom{\frac{A}{B}}b\right]} e^0_{n]}-e^0_{\left. \vphantom{\frac{A}{B}} b\right]} e^i_{n]})\mathfrak{D}_i\delta,
\ee
with antisymmetrisations assumed over $a\leftrightarrow b$ and $m \leftrightarrow n$, and with only two terms out of three in $\mathcal{G}$ having survived because the one having $e^{\nu}_d$ in it gets proportional to $\eta_{[ab]}=0$.
And it is very natural that we get this kind of uninviting term, since $\mathcal{F}_{ab}$ is indeed the generator of the local Lorentz transformations, while the second term in $\mathcal{C}_{mn}$ is not covariant with respect to such transformations. 

When computing \eqref{Ddterm} in terms of the smeared functions, and when it is summed with $\{\mathcal{G}_{ab}, \mathcal{F}_{mn}\}$, it means that we just have to replace the  $\mathfrak{D}^{(x_2)}_i\delta$ for the $\mathfrak{D}^{(x_1)}_i\delta$, and then the total contribution from the $\mathfrak{D} \delta$ part of $\{\mathcal{F}_{ab}, \mathcal{G}_{mn}\}$ and  $\{\mathcal{G}_{ab}, \mathcal{F}_{mn}\}$ gives 
\begin{multline} 
\label{additionP}
2\int dy\cdot \theta\left(\eta_{\left[\vphantom{\frac{A}{B}}a[m\right.}(e^i_{\left.\vphantom{\frac{A}{B}}b\right]} e^0_{n]}-e^0_{\left. \vphantom{\frac{A}{B}} b\right]} e^i_{n]})
\cdot C_{x_2}(y)\partial_i C_{x_1}(y)-\eta_{\left[\vphantom{\frac{A}{B}}m[a\right.}(e^i_{\left.\vphantom{\frac{A}{B}}n\right]} e^0_{b]}-e^0_{\left. \vphantom{\frac{A}{B}} n\right]} e^i_{b]})
\cdot C_{x_1}(y)\partial_i C_{x_2}(y)\right)\\
=2\int dy\cdot\theta\eta_{\left[\vphantom{\frac{A}{B}}a[m\right.}(e^i_{\left.\vphantom{\frac{A}{B}}b\right]} e^0_{n]}-e^0_{\left. \vphantom{\frac{A}{B}} b\right]} e^i_{n]})
\cdot \partial_i\left(\vphantom{\int} C_{x_1}(y)C_{x_2}(y)\right)
\end{multline}
being added to the simple Lorentz algebra part (\ref{newtermsP}), with the double copy of the algebra though. 

What is not seen directly from the derivatives of $\delta$-functionals, is that now we can now perform a new integration by parts in the last term we have got (\ref{additionP}), which brings us back to smearing with $C_{x_1}C_{x_2}$. Moreover, one can check that the derivative of the prefactor produces the $\mathcal{G}_{ab}$ expression again. Indeed, we have
$$2\partial_i \left(\theta (e^i_{a} e^0_{b} - e^i_{b} e^0_{a} )\right)=2\partial_i \left(\theta e^i_{[a} e^0_{b]} \right)=-2\theta \left( \partial_i \theta^{d}_{\nu} \right) \left( \e_{[a}^{0} \e_{b]}^{i} \e_{d}^{\nu} + \e_{[a}^{i} \e_{b]}^{\nu} \e_{d}^{0} + \e_{[a}^{\nu} \e_{b]}^{0} \e_{d}^{i} \right)=-\mathcal{G}_{ab}.$$
It turns the new potentially problematic contribution (\ref{additionP}) into
$$\int dy \cdot C_{x_1}C_{x_2}\cdot \eta_{\left[\vphantom{\frac{A}{B}}a[m\right.} \mathcal{G}_{\left.b\vphantom{\frac{A}{B}}\right]n]}=-\int dy \cdot C_{x_1}C_{x_2}\cdot(\eta_{an}{\mathcal G}_{bm} + \eta_{am}{\mathcal G}_{nb} + \eta_{bn}{\mathcal G}_{ma} + \eta_{bm}{\mathcal G}_{an})$$
and actually solves all the problems by removing one extra copy of Lorentz algebra expressions (\ref{newtermsP}) in terms of the non-covariant term and therefore closing the algebra:
\begin{equation} 
\label{badLor}
\{ {\mathcal C}_{ab}(x_1), {\mathcal C}_{mn}(x_2)\}
= (\eta_{an}{\mathcal C}_{bm} + \eta_{am}{\mathcal C}_{nb} + \eta_{bn}{\mathcal C}_{ma} + \eta_{bm}{\mathcal C}_{an})\cdot \delta(x_1-x_2).
\end{equation}

To be more precise, the formula (\ref{badLor}) is not fully correct. It will be better to denote the smeared constraints as $\mathfrak{C}_{ab}[C_x]=\int dy C_x(y) \mathcal{C}_{ab}(y)$ and write it as
$$\{ {\mathfrak C}_{ab}[C_{x_1}], {\mathfrak C}_{mn}[C_{x_2}]\}
= \eta_{an}{\mathfrak C}_{bm}[C_{x_1}C_{x_2}] + \eta_{am}{\mathfrak C}_{nb}[C_{x_1}C_{x_2}] + \eta_{bn}{\mathfrak C}_{ma}[C_{x_1}C_{x_2}] + \eta_{bm}{\mathfrak C}_{an}[C_{x_1}C_{x_2}].$$
Indeed, the formula (\ref{badLor}) is not what we get at the level of operators themselves. The actual expression of the singular bracket as an operator is
\begin{multline*}
\{ {\mathcal F}_{ab}, {\mathcal G}_{mn}\}+\{\mathcal{G}_{ab} , \mathcal{F}_{mn}\}= 2\eta_{\left[\vphantom{\frac{A}{B}}a[n\right.} \mathcal{G}_{\left.b\vphantom{\frac{A}{B}}\right]m]}\cdot \delta+ 2\theta\eta_{\left[\vphantom{\frac{A}{B}}a[m\right.}(e^i_{\left.\vphantom{\frac{A}{B}}b\right]} e^0_{n]}-e^0_{\left. \vphantom{\frac{A}{B}} b\right]} e^i_{n]})\cdot (\mathfrak{D}^{(x_1)}_i\delta +\mathfrak{D}^{(x_2)}_i\delta)\\
=\eta_{\left[\vphantom{\frac{A}{B}}a[n\right.} \mathcal{G}_{\left.b\vphantom{\frac{A}{B}}\right]m]}\cdot \delta+(\mathfrak{D}^{(x_1)}_i +\mathfrak{D}^{(x_2)}_i)\cdot 2\theta\eta_{\left[\vphantom{\frac{A}{B}}a[m\right.}(e^i_{\left.\vphantom{\frac{A}{B}}b\right]} e^0_{n]}-e^0_{\left. \vphantom{\frac{A}{B}} b\right]} e^i_{n]}) \ \delta.
\end{multline*}
In other words, it got to have an extra term being an operator which first makes two arguments of a test function $\mathcal{C}(z_1,z_2)$ equal to each other, multiplies it by some another function of the same variable, and then takes the derivative with respect to it. It is a full derivative, not just with respect to those arguments which had been some particular one of the $z_1$ and $z_2$ before. Therefore, upon integration this new term goes away, but only upon integration, and only if we have not multiplied the result by some another function before. This is why the extra integration by parts worked well with the formula (\ref{additionP}). Therefore, the smeared quantities make up the closed Lorentz algebra indeed, but the operators of the singular brackets of local quantities do also have another term which adds and extra function to the results of acting on a test function, which is a full derivative though.

Once again, this closure of the algebra (\ref{badLor}) is not something what directly comes as a limit of operators. We must perform integration of the result to get it. It works because the mismatch is an operator which gives a total derivative function when acting on an arbitrary test function, and therefore it becomes zero upon integration. On one hand, it is not surprising since the symmetry of the action was also only due to integration. On the other hand, it complicates the Hamiltonian calculations and might potentially be problematic. In case of performing canonical quantisation, it seems to become a certain problem with commutators making the symmetry anomalous. Though other methods of quantisation would probably see these effects directly via the non-invariant boundary term of the action.

At the same time, allowing ourselves to take the algebra simply as the usual Lorentz one (\ref{badLor}) makes the Leibnitz rule for Poisson brackets not applicable any longer, not even for some extra factors not depending on any tetrad components at all. Any other function which got to multiply $\mathcal{G}_{mn}$ in the term with the $\mathfrak{D}^{(x_1)}_i +\mathfrak{D}^{(x_2)}_i$ operator would also get differentiated upon the extra integration by parts. In particular, in the case of the Jordan frame representation of $f(T)$ gravity where the Lagrangian is written as $\phi\mathbb T + V(\phi)$, the auxiliary scalar field $\phi$ comes always with the smearing function of the $\mathcal G$ part. It effectively corresponds to smearing (\ref{additionP}) with $\phi(y)\partial_i\left(\vphantom{\int} C_{x_1}(y)C_{x_2}(y)\right)$ instead of just $\partial_i\left(\vphantom{\int} C_{x_1}(y)C_{x_2}(y)\right)$, and upon integration by parts we get violation of Lorentz algebra with terms proportional to $\theta \eta_{am}(e^i_{b} e^0_{n}-e^0_{b} e^i_{n}))\partial_i\phi$. Good news is that the Poisson brackets, though not forming a closed algebra, are proportional to the universal $\delta(x_1-x_2)$ without more complicated structures (admitting the trick of this extra integration by parts). This Lorentz violation at the level of the primary constraints has been reported in  \cite{Li:2011rn,Blagojevic:2020dyq}, although an important remark is that such violation depends on two factors with different physical meanings. One of the factors is $\partial_i\phi$, which vanishes when $\mathbb T$ is  constant or only dependent on time, fully recovering the algebra; this fact has been pointed out in \cite{Blagojevic:2020dyq}. However, the antisymmetric combination of tetrads $(e^i_{b} e^0_{n}-e^0_{b} e^i_{n})$ can also achieve an ``accidental gauge restoration'', as this quantity can vanish for particular tetrad configurations.

\subsection{Poisson brackets with $\Pi^{0}_{a}$ constraints}

After having considered the Lorentz algebra, recall that we also have another set of primary constraints, $\Phi^0_a=\Pi^0_a$. They obviously commute with the usual Lorentz algebra,
$$\{\mathcal{F}_{ab}\ ,  \ \Pi^0_c\}=\Pi^0_{[a}\eta_{b]c}.$$
To find the property 
\begin{equation}
\label{moreconst}
\{\mathcal{C}_{ab}\ ,  \ \Pi^0_c\}=\Pi^0_{[a}\eta_{b]c},
\end{equation}
which is important in TEGR and in its nonlinear modifications, we need to check that $\{\mathcal{G}_{ab}\ ,  \ \Pi^0_c\}=0$. 

In principle, this task has no complications of spatial derivatives, since the gradient $\partial_i\theta^c_{\nu}$ enters the constraint in such a way that summation over $\nu$ can be restricted to spatial only with no change at all. However, if not being afraid of the derivatives, it is even easier to use the fact we found in the previous subsection that $\mathcal{G}_{ab}=-2\partial_i \left(\theta e^i_{[a} e^0_{b]} \right)$. A simple calculation then shows that $\frac{\delta}{\delta\theta^c_0}\left(\theta e^i_{[a} e^0_{b]} \right)=0$ which finishes the proof. Indeed, variations of the two lateral factors give terms $\theta e^i_{[a}e^0_{b]}e^0_c$ with opposite signs which cancel each other, while variation of the middle one gives $\theta e^0_{[a}e^0_{b]}e^i_c=0$.

\section{Momenta in covariant TEGR and pure-tetrad constraints}
\label{sec:covariant}

In this section we will present the covariant formulation of teleparallel gravity \cite{Golovnev:2017dox,Krssak:2018ywd}, and derive the associated primary constraints and their algebra. Teleparallel gravity is endowed with a gauge symmetry under simultaneous transformations of both the tetrad and the spin connection
\be 
\E^{a}_{\mu} \longrightarrow \Lambda^{a}_{b} \E^{b}_{\mu}, \hspace{8mm} \omega^{a}_{\hphantom{a}b\mu} \longrightarrow 
\omega^{a}{}_{b \mu} = \Lambda^{a}_{c} \omega^{c}{}_{d\mu} (\Lambda^{-1})^{d}_{b} -  (\Lambda^{-1})^{c}_{b} \partial_{\mu} \Lambda^{a}_{c},
\ee 
where $\Lambda^{a}_{b}$ is a matrix representing a Lorentz transformation. The transformed connection is also metric compatible, and vanishes the Riemann tensor \eqref{Riemann}, therefore keeping up with a teleparallel spacetime. This simultaneous transformation introduces an additional freedom not requiring to fix the spin connection to the Weitzenb\"{o}ck gauge, in other words, Lorentz invariance is recovered at the cost of introducing additional fields $\Lambda$. 

Therefore, when Lorentz transforming the Weitzenb\"{o}ck connection, one gets a nonvanishing flat spin connection
\be 
\omega^{a}{}_{b \mu} =- (\Lambda^{-1})^{c}_{b} \partial_{\mu} \Lambda^{a}_{c} 
\label{inertial}
\ee 
which has been dubbed the ``inertial'' spin connection. This is the most general form of a spin connection which can be obtained by Lorentz-transforming the zero spin connection of the pure tetrad model, and this way of presenting an arbitrary flat metric-compatible spin connection is always possible, 
at least modulo possible global obstacles for that.
Consequently, the Lorentz-covariant formulation of any teleparallel theory of gravity can be obtained by the replacement
\be 
\partial_{\mu}\theta^a_{\nu}
\longrightarrow {\mathcal D}_{\mu} \theta^a_{\nu} \equiv \partial_{\mu} \theta^{a}_{\nu} - (\Lambda^{-1})^{c}_{b} (\partial_i \Lambda^{a}_{c})\theta^b_{\nu}
\ee
at the level of the action \eqref{Lpremetric}. Therefore, the same substitution applies also to the definition of tetrad momenta \eqref{LorentzIntermediate} and the constraints.

\subsection{Covariant pure-tetrad primary constraints}

Since the changes from passing to the covariant version of the theory are very simple, it is straightforward to see that the trivial primary constraints $\Phi^0_a=\Pi^0_a=0$ remain the same, while the old pure-tetrad Lorentz constraints acquire the form of
\be 
\label{covptc}
C^{\mathrm{cov}}_{ab} = \Pi_{[a}^{\mu}\eta_{b]c} \theta^{c}_{\mu} - 2 \theta \left( {\mathcal D}_{\mu}\theta^c_{\nu} \right) \left( \e_{[a}^{0} \e_{b]}^{i} \e_{c}^{\nu} + \e_{[a}^{i} \e_{b]}^{\nu} \e_{c}^{0} + \e_{[a}^{\nu} \e_{b]}^{0} \e_{c}^{i} \right),
\ee 
which can be compared with the non-covariant version of the constraint by the addition of a new term:
\begin{eqnarray*}
C^{\mathrm{cov}}_{ab} & = & C_{ab}+ 2 \theta (\Lambda^{-1})^{d}_{f} (\partial_i \Lambda^{c}_{d})\theta^f_{\nu} \left( \e_{[a}^{0} \e_{b]}^{i} \e_{c}^{\nu} + \e_{[a}^{i} \e_{b]}^{\nu} \e_{c}^{0} + \e_{[a}^{\nu} \e_{b]}^{0} \e_{c}^{i} \right)\\
& = & C_{ab}+ 2 \theta\left( (\Lambda^{-1})^{d}_{[b} (\partial_i \Lambda^{c}_{d})  \e_{a]}^{i}  \e_{c}^{0} +(\Lambda^{-1})^{d}_{[a} (\partial_i \Lambda^{c}_{d})  \e_{b]}^{0} \e_{c}^{i} \right)\\
& = & C_{ab}+ 2 \theta  (\Lambda^{-1})^{d}_{[a} (\partial_i \Lambda^{c}_{d}) \left( \e_{b]}^{0} \e_{c}^{i}-\e_{b]}^{i} \e_{c}^{0} \right).
\end{eqnarray*}
Let us denote this new addition as
\be 
\mathcal{H}_{ab}=2 \theta  (\Lambda^{-1})^{d}_{[a} (\partial_i \Lambda^{c}_{d}) \left( \e_{b]}^{0} \e_{c}^{i}-\e_{b]}^{i} \e_{c}^{0} \right).
\ee 
Note that it is antisymmetrised with respect to $b \leftrightarrow c$, and therefore, as we have seen before, it also has zero Poisson brackets with the other constraints $\Phi^0_a=\Pi^0_a$, the same as the previous part $\mathcal{G}_{ab}$.

In order to check the Lorentz algebra, we need to evaluate the new Poisson bracket:
\begin{eqnarray*}
\{\mathcal{F}_{ab},\mathcal{H}_{mn}\} & = & \left\{ \Pi_{[a}^{\mu}\eta_{b]c} \theta^{c}_{\mu}\ ,\ 2 \theta  (\Lambda^{-1})^{e}_{[m} (\partial_i \Lambda^{d}_{e}) \left( \e_{n]}^{0} \e_{d}^{i}-\e_{n]}^{i} \e_{d}^{0} \right)\right\} \\
& = & 2\theta\eta_{\left[\vphantom{\frac{A}{B}} b[n\right.}(\Lambda^{-1})^{e}_{m]} (\partial_i \Lambda^{d}_{e}) (e^0_{\left.a\vphantom{\frac{A}{B}}\right]} e^i_d- e^i_{\left.a\vphantom{\frac{A}{B}}\right]} e^0_d) +2\theta\eta_{\left[\vphantom{\frac{A}{B}} bd\right.}
(\Lambda^{-1})^{e}_{[m} (\partial_i \Lambda^{d}_{e})
(e^0_{n]} e^i_{\left. a 
\vphantom{\frac{A}{B}} 
\right] }- e^i_{n]} e^0_{\left. a \vphantom{\frac{A}{B}} 
\right]}).
\end{eqnarray*}
The first term is one half of the normal Lorentz algebra, $-\eta_{\left[\vphantom{\frac{A}{B}}a[m\right. }\mathcal{H}_{\left.b\vphantom{\frac{A}{B}}\right]n]}=\eta_{\left[\vphantom{\frac{A}{B}}b[n\right. }\mathcal{H}_{\left.m\vphantom{\frac{A}{B}}]a\right]}$, what it lacks is $m \leftrightarrow a$ antisymmetrisation to be expressed indeed in terms of $\mathcal{H}_{ma}$. The reason is that one of the two defining indices belongs to the $\Lambda$-part which does not get transformed by the generator $\mathcal F$. The second term is an extra, unwanted contribution, due to a dumb index $d$. It would have been cancelled if the spin connection part had been transformed, too.

This result is not surprising, as the old pure-tetrad constraint still has only the part which rotates the tetrad, but not the $\Lambda$ matrix, therefore when we calculate the Poisson brackets of these constraints with themselves, we get in the $\{\mathcal{F}_{ab},\mathcal{H}_{mn}\}$ terms the transformation terms related with the indices $n$ and $d$, and with $m$ and $d$ separately, instead of just $m$ and $n$ everywhere. Once we have introduced the spin connection in terms of the Lorentz matrices, the theory fails to be covariant even under the global Lorentz transformations, if applied to the tetrad only. However, it is still ``pseudo-invariant''. Therefore we must have a look at the full bracket.

To find the full bracket we need to also add $\{\mathcal{H}_{ab},\mathcal{F}_{mn}\}$ to the above result. Then it is easy to see that the first terms combine to produce the full Lorentz algebra of $\eta_{\left[\vphantom{\frac{A}{B}}b[n\right. }\mathcal{H}_{\left.m\vphantom{\frac{A}{B}}]a\right]}$. At the same time, the second terms go as
\begin{multline*}
2\theta\eta_{\left[\vphantom{\frac{A}{B}} bd\right.}
(\Lambda^{-1})^{e}_{[m} (\partial_i \Lambda^{d}_{e})
(e^0_{n]} e^i_{\left. a 
\vphantom{\frac{A}{B}} 
\right] }- e^i_{n]} e^0_{\left. a \vphantom{\frac{A}{B}} 
\right]})-2\theta\eta_{\left[\vphantom{\frac{A}{B}} nd\right.}
(\Lambda^{-1})^{e}_{[a} (\partial_i \Lambda^{d}_{e})
(e^0_{n]} e^i_{\left. b 
\vphantom{\frac{A}{B}} 
\right] }- e^i_{b]} e^0_{\left. m \vphantom{\frac{A}{B}} 
\right]})\\
=2\theta\left(\eta_{\left[\vphantom{\frac{A}{B}} bd\right.}
(\Lambda^{-1})^{e}_{[m} (\partial_i \Lambda^{d}_{e})+\eta_{\left[\vphantom{\frac{A}{B}} md\right.}
(\Lambda^{-1})^{e}_{[b} (\partial_i \Lambda^{d}_{e})\right)
(e^0_{n]} e^i_{\left. a 
\vphantom{\frac{A}{B}} 
\right] }- e^i_{n]} e^0_{\left. a \vphantom{\frac{A}{B}} 
\right]})=0,
\end{multline*}
due to antisymmetry of the spin connection. Altogether, we finally have the proper Lorentz algebra again: $$\{\mathcal{C}^{\mathrm{cov}}_{ab},\mathcal{C}^{\mathrm{cov}}_{mn}\}=\eta_{\left[\vphantom{\frac{A}{B}}b[n\right. }\mathcal{C}^{\mathrm{cov}}_{\left.m\vphantom{\frac{A}{B}}]a\right]}.$$

Note that covariantisation did not solve any problem. This algebra is closed again due to the extra integration by parts for the terms coming from the $\partial_i \theta^a_{\mu}$ adverse term. However, the new addition kept the Lorentz algebra satisfied without any new trouble since it did not include any spatial derivatives of fields which have their momenta in the constraint.

\subsection{Additional momenta of Lorentz matrices}

In the following we derive the primary constraints associated with the incorporation of the covariant setup. The derivation we are going to perform differs from the one made in \cite{Blixt:2018znp,Blixt:2019mkt}, because we are not making use of auxiliary variables to impose antisymmetry on the canonical momenta. In fact, it is not accurate to call the antisymmetric quantities obtained in \cite{Blixt:2018znp} momenta, since they are given in terms of auxiliary variables instead of the velocities of Lorentz matrices properly. Instead, we vary in terms of the velocities $\dot{\Lambda}^a_b$, therefore getting indeed a definition for the corresponding momenta $P^b_a$. 

The Lorentzian momenta come as
\be 
P^b_a= \frac{\delta\mathcal L}{\delta\dot \Lambda^a_b} =\frac{\delta\mathcal L}{\delta{\mathcal D}_0  \theta^c_{\mu}}\cdot \frac{\delta{\mathcal D}_0 \theta^c_{\mu}}{\delta\dot \Lambda^a_b} =-\Pi^{\mu}_a\theta^c_{\mu}(\Lambda^{-1})^b_c
\ee 
which in comparison with the momenta for the tetrad
\be 
\Pi^{\mu}_a=\frac{\delta\mathcal L}{\delta\dot \theta^a_{\mu}}=\frac{\delta\mathcal L}{\delta{\mathcal D}_0 \theta^a_{\mu}},
\ee 
implies a linear relation among the two kinds of momenta
\begin{equation} 
\label{momrel}
P^b_a + \Pi^{\mu}_a \theta^c_{\mu}(\Lambda^{-1})^b_c=0.
\end{equation}
This relation is a legit primary constraint, but certain caution has to be taken. This is because we are taking the matrices $\Lambda$ fully arbitrary with 16 independent components, which are not completely independent of each other if we want them to represent proper Lorentz transformations. A direct but rather cumbersome way around of this redundancy would be to either explicitly parametrise the Lorentz group by 6 parameters or to introduce constraints $\Lambda^a_c \eta_{ab} \Lambda^b_d=\eta_{cd}$ defining the Lorentz group via Lagrange multiplier terms in the action.

To avoid unnecessary complications, we can go using the symmetry properties. Recall that, analogously, the momenta of spatial metric in ADM formalism of GR are symmetrised, and this is taken into account in the Poisson brackets. Indeed, the symmetry of the metric requires $g_{ij}=g_{ji}$, but it would be very hard to really treat them as one and the same variable ignoring the spatially covariant way of writing. If we formally take them independently, one has to treat them symmetrically also in deriving the momenta to not potentially break the symmetry of the metric by the Hamiltonian evolution.

Even though the spatial metric momentum would automatically go out symmetric  from any covariant expression, it can be even formally imposed by taking the momenta as $\pi^{ij}\to \frac12 (\pi^{ij}+\pi^{ji})$, which transforms the trivial Poisson bracket $\{g_{ij},\pi^{kl}\}=\delta^{k}_{i}\delta^{l}_{j}$ into the one which works correctly for this formalism: $\{g_{ij},\pi^{kl}\}=\frac12 \left(\delta^{k}_{i}\delta^{l}_{j}+\delta^{l}_{i}\delta^{k}_{j}\right)$. When we treat different metric components as independent, the Poisson bracket gives $1$ for diagonal elements with their corresponding momenta, and $1/2$ for non-diagonal ones with each one of the two corresponding momenta representing one half of the real quantity. One can also view the same result by changing $\pi^{ij}$ in all expressions for ${\tilde\pi}^{ij}= \frac12 (\pi^{ij}+\pi^{ji})$ with $\{g_{ij},{\tilde \pi}^{kl}\}=\frac12 \left(\delta^{k}_{i}\delta^{l}_{j}+\delta^{l}_{i}\delta^{k}_{j}\right)$ while retaining the standard fundamental bracket $\{g_{ij},\pi^{kl}\}=\delta^{k}_{i}\delta^{l}_{j}$.

\subsubsection{The symmetry property of Lorentzian momenta}

Our case of Lorentz matrices is more subtle since their symmetry properties are not that elementary. The antisymmetry is present in the elements of the corresponding Lie algebra. One could indeed incorporate the Lie group element in terms of the Lie algebra into the action, by defining the Lorentz matrix $\Lambda=e^{\lambda}$ with antisymmetric matrix $\lambda_{ab}=\eta_{ac}\lambda^c_b$. Then the momentum of $\lambda$ would be antisymmetric, however working with explicit derivatives of matrix exponentials would be an unpleasant task, for the velocity then enters the action via
$$\frac{d}{dt}e^{\lambda}=\dot\lambda+\frac12(\dot\lambda \lambda+\lambda \dot\lambda)+\frac16 (\dot\lambda \lambda^2+\lambda \dot\lambda \lambda+\lambda^2 \dot\lambda)+\ldots,$$
and the $\lambda$ and $\dot\lambda$ matrices in general do not commute.

In other words, the subtlety of working with momenta of $\Lambda$ is in the fact that these matrices are not antisymmetric by themselves but are related to antisymmetry. Indeed, an infinitesimal variation of $\Lambda$ belongs to the corresponding Lie algebra of matrices $\lambda$ with antisymmetric $\eta\lambda$ only if taken around the unit element of the Lie group, $\Lambda=I$. The Lie algebra is the tangent space of the Lie group at unity, however the group structure allows us to obtain all the properties around a different point on the group via the group transformation from the unity. And then, the variation can be transformed from being around $I$ to being around $\Lambda$ by an action of this very matrix:
$$\delta \Lambda = \Lambda \lambda.$$

Indeed, let us write an infinitesimal one-parameter family of Lorentz matrices as $\Lambda(t)=\Lambda(0)+\delta \Lambda(t)$, and act with $\Lambda^{-1}(0)$ upon that. Then the matrix $\Lambda^{-1}(0)\cdot\Lambda(t)$ is infinitesimally close to unity $I$ and is given by $I+\Lambda^{-1}(0)\cdot\delta \Lambda(t)$. We conclude that the matrix $\Lambda^{-1}(0)\cdot\delta \Lambda(t)$ with infinitesimal $\delta\Lambda$ belongs to the tangent space of the group at unity $I$, which is the Lie algebra of the Lorentz group. Therefore, a smooth infinitesimal family of matrices $\Lambda(t)$ can be written as 
\begin{equation}
\label{Lfamily}
\Lambda(t)\approx\Lambda(0)\cdot (I+ \lambda(t))
\end{equation}
with $\lambda(t)$ belonging to the Lie algebra, i.e. $\lambda_{ab}=-\lambda_{ba}$. 

Note that the $\lambda$ matrix in the formula (\ref{Lfamily}) is of course not the same as in $\Lambda(t)=e^{\lambda(t)}$. The important point is just that it also belongs to the Lie algebra, and therefore is antisymmetric. Then we can write for the time derivative 
\begin{equation}
\label{Lder}
\dot\Lambda^a_b=\Lambda^a_c\dot\lambda^c_b.
\end{equation}
In other words, a derivative of $\Lambda$ can be expressed in terms of multiplying $\Lambda$ by some  antisymmetric matrix (if to raise its free index, or in Euclidean signature).
As an example, one can take a two-dimensional rotation matrix and calculate its derivative:
$$\frac{\partial}{\partial\phi} \left(
\begin{array}{cc}
\cos\phi & \sin\phi  \\
-\sin\phi &  \cos\phi 
\end{array}
\right)=
\left(
\begin{array}{cc}
-\sin\phi & \cos\phi  \\
-\cos\phi &  -\sin\phi 
\end{array}
\right)=
\left(
\begin{array}{cc}
\cos\phi & \sin\phi  \\
-\sin\phi &  \cos\phi 
\end{array}
\right)
\cdot
\left(
\begin{array}{cc}
0 & 1  \\
-1 &  0
\end{array}
\right).$$
Note that the result (\ref{Lder}) is just a very important and well-known property of covariant teleparallel gravity, it is nothing but antisymmetry of $\Lambda^{-1}\partial\Lambda$ which is the spin connection components. 

Now we are ready to discuss the momenta. When we take a derivative of the Lagrangian with respect to the velocity $\dot\Lambda$ as $\delta{\mathcal L}=P^b_a\delta \dot\Lambda^a_b$, it all means that $P^c_a\Lambda_{c}^{b}$ is antisymmetric:
\begin{equation}
\label{symmP}
P_a^c\Lambda_{c}^d\eta_{db}=-P_b^c\Lambda_{c}^d\eta_{da}.
\end{equation}

\subsubsection{Alternative derivation of the symmetry}

Another derivation of the symmetry property (\ref{symmP}) of $P$ can go as follows. Even though it is not so simple to differentiate a matrix exponential, we can start by taking a look at an equation 
$$e^X e^Y =e^Z$$ 
for matrices $X$, $Y$ and $Z$. On the left hand side we have 
\begin{equation} 
\label{exprod}
e^X e^Y =I+X+Y+\frac12 X^2+\frac12 Y^2+XY + \ldots.
\end{equation}
Then we see that, at the linear level $Z=X+Y$, which would be the full answer if $X$ and $Y$ commuted with each other, but generically it gives 
$$\frac12 Z^2=\frac12(X+Y)^2=\frac12 (X^2+ Y^2+XY+YX)+\ldots$$
in the second order of $e^Z$, and therefore, comparing with the equation (\ref{exprod}), we have to correct the sum of the two matrices by their commutator:
$$Z=X+Y+\frac12[X,Y]+ \mathrm{higher\ order\ commutators}.$$
Some playing with combinatorics can explicitly give the full series, which is known as Baker-Campbell-Hausdorf formula. However, the important point for us is just that the result goes solely in terms of commutators, since it is precisely them being non-zero what prevents the $X+Y$ matrix from being the full answer.

Now, let us specify our example to an infinitesimal case, namely 
\be 
e^{-X}e^{X+\delta X}=e^Z,
\ee
which gives 
$$Z=\delta X-\frac12[X,\delta X]+\ldots.$$ 
Taking $X\equiv\lambda$ and leaving only linear in $\delta X\equiv \dot\lambda \delta t$ terms on the right hand side, $e^Z\approx I+Z$, we can get
$$e^{-\lambda}\frac{d}{dt}e^{\lambda}=\dot\lambda-\frac12[\lambda,\dot\lambda]+\mathrm{higher\ order\ commutators\ of\ one\ } \dot\lambda \mathrm{\ with\ many\ } \lambda$$
which obviously belongs to the Lie algebra, since the latter is closed under linear combinations and commutations. Belonging to the Lie algebra is the antisymmetry property. Of course, it is again just another way to see the antisymmetry of the spin connection given by $\Lambda^{-1}\partial\Lambda$.

\subsubsection{Implications of the symmetry}

If all the components of $\Lambda$ were independent, we would define the Poisson bracket as $\{\Lambda^d_c,P^a_b\}=\delta^a_c\delta^d_b$. In case of the metric components we can take an analogous bracket and get the correct one by $\pi^{ij}\to \frac12 (\pi^{ij}+\pi^{ji})$. Now we have to impose instead that 
\begin{equation}
\label{reqP}
P_a^c\Lambda_{c}^{d}\eta_{db}+P_b^c\Lambda_{c}^{d}\eta_{da}=0.
\end{equation}
This is less trivial than simple (anti)symmetrisation.

Therefore, let us present it as changing the momentum $P^{ab}$ for 
$${\tilde P}^a_b=\frac12 \left( P^a_b-(\Lambda^{-1})^a_g \eta^{gd}P^c_d \Lambda^f_c\eta_{fb}\right)=\frac12 \left(\delta^a_c \delta^d_b-(\Lambda^{-1})^a_g \eta^{gd} \Lambda^f_c\eta_{fb}\right)P^c_d.$$
It gives us new Poisson brackets
$$\{\Lambda^d_c,{\tilde P}^a_b\}=\frac12 \left(\delta^a_c \delta^d_b-(\Lambda^{-1})^a_g \eta^{gd} \Lambda^f_c\eta_{fb}\right),$$
and even $\{{\tilde P}^c_d,{\tilde P}^a_b\}\neq 0$.

If we do calculations having only antisymmetrised $P\Lambda$ combinations in all the expressions, then it is not necessary to switch to $\tilde P$ momenta since anyway 
$${\tilde P}_a^c\Lambda^d_c\eta_{db}-{\tilde P}_b^c\Lambda^d_c\eta_{da}={P}_a^c\Lambda^d_c\eta_{db}-{P}_b^c\Lambda^d_c\eta_{da},$$ 
and this mere antisymmetrisation suffices, still safely keeping up with the trivial basic brackets $\{\Lambda_c^d,P^a_b\}=\delta^a_c\delta^d_b$ and $\{P^a_b, P^c_d\}=0$. 

Note also that one of the effects of taking these symmetry properties into account explicitly is to not violate the Lorentz property $\Lambda^a_c \eta_{ab} \Lambda^b_d=\eta_{cd}$. And we can easily check indeed that
$$\{{\tilde P}^a_b\ ,\ \Lambda^e_c \eta_{ef} \Lambda^f_d\}=\frac12 \{ \left(\delta^a_c \delta^d_b-(\Lambda^{-1})^a_g \eta^{gd} \Lambda^f_c\eta_{fb}\right)P^c_d\ ,\ \Lambda^k_m \eta_{kp} \Lambda^p_n\}=0.$$
It is similar to the fact that $\{{\tilde \pi}^{ij}, g_{ij}-g_{ji}\}=0$.

\section{Primary constraints of simultaneous Lorentz transformations}
\label{sec:simult}

Finally, combining the relation (\ref{momrel}) between the tetrad and Lorentz momenta with the symmetry property (\ref{symmP}) of the latter, we find the constraints corresponding to the simultaneous Lorentz transformation:
\be 
{\mathcal C}^{\prime}_{ab} = \Pi^{\mu}_{[a}\eta_{b]c} \theta^c_{\mu} + {\tilde P}^c_{[a}\eta_{b]d}\Lambda^d_c = \Pi^{\mu}_a\theta^c_{\mu}\eta_{cb} -\Pi^{\mu}_b\theta^c_{\mu}\eta_{ca} + { P}^c_a\Lambda^d_c\eta_{db} -  { P}^c_b\Lambda^d_c\eta_{da} 
\ee
by simply restricting the relation (\ref{momrel}) to its physically meaningful, non-redundant part. They clearly form the correct Lie algebra of the Lorentz group
\be 
\{ {\mathcal C}^{\prime}_{ab}, {\mathcal C}^{\prime}_{cd}\}
= \eta_{ad}{\mathcal C}^{\prime}_{bc} + \eta_{ac}{\mathcal C}^{\prime}_{db} + \eta_{bd}{\mathcal C}^{\prime}_{ca} + \eta_{bc}{\mathcal C}^{\prime}_{ad}=-\eta_{\left[\vphantom{\frac{A}{B}}a[c\right. }{\mathcal C}^{\prime}_{\left.b\vphantom{\frac{A}{B}}\right]d]},
\ee
with the usual simple laws of transformation for tensors:
\begin{eqnarray}
\{P_{[a}^c \eta_{b]d} \Lambda^d_c \ , \ \Lambda^m_n\} & = & -\delta^m_{[a}\eta_{b]d}\Lambda^d_n ,\\
 \{P_{[a}^c\eta_{b]d} \Lambda^d_c \ , \ (\Lambda^{-1})^m_n\}  & = & (\Lambda^{-1})^m_{[a}\eta_{nb]},
\end{eqnarray}
and with basically the same problematic piece as in the case of the tetrad when going for brackets with the old constraints:
$$\{P_{[a}^c(x_1)\eta_{b]d} \Lambda^d_c(x_1) \ , \ \partial_i\Lambda^m_n(x_2)\} = -\delta^m_{[a} \eta_{b]d} \left((\partial_i\Lambda^d_n)\cdot\delta(x_1-x_2) +\Lambda^d_n\cdot\mathcal{D}^{(x_1)}_i\delta(x_1-x_2)\right).$$

Using the commutator with the spatial derivative, it is not difficult to see that the new constraints are first class, also with respect to the covariantised pure-tetrad constraints \eqref{covptc}. The point is that the $\Pi$ and $P$ parts together transform the Lorentz-covariant spatial derivatives $\mathcal{D}_{\mu}\theta^a_{\nu}$ correctly, with the $\mathfrak{D}\delta$ parts cancelling each other:
$$\{\Pi_{[a}^{\mu}\eta_{b]c} \theta^{c}_{\mu} + P_{[a}^c  \eta_{b]d} \Lambda^d_c\ , \ \partial_i \theta^{m}_{\lambda} - (\Lambda^{-1})^{e}_{f} \partial_i \Lambda^{m}_{e}\theta^f_{\lambda}\}=-\delta^m_{[a} \eta_{db]} \left(\partial_i \theta^{d}_{\lambda} - (\Lambda^{-1})^{e}_{f} \partial_i \Lambda^{d}_{e}\theta^f_{\lambda}\right)\cdot\delta(x_1-x_2).$$

Let us more conveniently illustrate it at the level of more elementary quantities
\be 
\mathfrak{D}^{c}_{di} \equiv e^{\lambda}_{d} \mathcal{D}_i \theta^{c}_{\lambda}= e^{\lambda}_{d} \partial_i \theta^{c}_{\lambda} - (\Lambda^{-1})^{e}_{d} \partial_i \Lambda^{c}_{e} 
\ee
in terms of which the covariantised old Lorentz constraint can be written down as
$$\mathcal{C}^{\mathrm{cov}}_{ab} =  \Pi^{\mu}_{[a}\eta_{b]e} \E^{e}_{\mu} - 2\theta ( \e^{0}_{[a} \e^{i}_{b]} \mathfrak{D}^{c}_{ci} + \e^{i}_{[a} \mathfrak{D}^{c}_{b]i} e^{0}_{c]} + \mathfrak{D}^{c}_{[ai} \e^{0}_{b]} \e^{i}_{c} ).$$
To calculate the Poisson bracket of the two constraints, let us evaluate a simpler bracket: 
\begin{eqnarray*}
\{ \mathcal{C}^{\prime}_{ab}, \mathfrak{D}^{m}{}_{ni} \} & = &  
\{ \Pi^{\mu}_{[a}\eta_{b]c} \E^{c}_{\mu}, e^{\lambda}_{n} \partial_i \theta^{m}_{\lambda}  \}-
\{ P_{[a}^c\eta_{b]d} \Lambda^d_c , (\Lambda^{-1})^{f}_{n} \partial_i \Lambda^{m}_{f} \}\\
& = & \left(e^{\lambda}_{[a}\eta_{b]n}\partial_i\theta^m_{\lambda}-e^{\lambda}_{n}\delta^m_{[a}\eta_{b]k}\partial_i\theta^k_{\lambda}\right)\cdot\delta-\delta^m_{[a}\eta_{b]n}\cdot\mathcal{D}_i\delta \\ 
& & - \left((\Lambda^{-1})^{f}_{[a}\eta_{b]n}\partial_i\Lambda^m_f-(\Lambda^{-1})^f_{n}\delta^m_{[a}\eta_{b]k}\partial_i\Lambda^k_f\right)\cdot\delta +\delta^m_{[a}\eta_{b]n}\cdot\mathcal{D}_i\delta\\
& = & \left( \eta_{d[n} \mathfrak{D}^{m}{}_{a]i} -\eta_{k[b}\delta^m_{a]} \mathfrak{D}^{k}{}_{ni}\right) \cdot \delta(x_1-x_2)
\end{eqnarray*}
which is the perfectly covariant transformation law related to the two Latin indices of $\mathfrak D$.
After that the calculations are quite elementary and yield
$$\{ {\mathcal C}^{\prime}_{ab}, {\mathcal C}^{\mathrm{cov}}_{cd}\}
=-\eta_{\left[\vphantom{\frac{A}{B}}a[c\right. }{\mathcal C}^{\mathrm{cov}}_{\left.b\vphantom{\frac{A}{B}}\right]d]}.$$ 
This is a very natural result. This is how every locally Lorentz invariant quantity must behave under the transformations generated by ${\mathcal C}^{\prime}_{ab}$. Note that for this formula we don't need to perform the trick with an extra integration by parts, the $\mathcal{D}_i$ parts cancel each other as operators.

Finally, let us claim the last point of studying the algebra of the primary constraints,
$$\{\mathcal{C}^{\prime}_{ab}\ ,  \ \Pi^0_c\}=\{\mathcal{C}^{\mathrm{cov}}_{ab}\ ,  \ \Pi^0_c\}=\{\mathcal{C}_{ab}\ ,  \ \Pi^0_c\}=\Pi^0_{[a}\eta_{b]c}$$
which should be obvious by now.

\section{Conclusions}
\label{sec:concl}

In this article we have exposed with great detail the definitions of Lorentz symmetries in teleparallel theories of gravity through the lens of constrained Hamiltonian mechanics. Our main findings show that the primary constraints coming from simultaneous Lorentz transformations of tetrad and spin connection form the algebra of Lorentz generators, generate infintesimal Lorentz transformations on the tetrad, and fully commute with Lorentz constraints associated with Lorentz rotations of only the tetrad.

In order to make our exposition more comprehensible to the reader, we first study non-covariant teleparallel gravity, that is, the theory whose Lagrangian considers only the tetrad as a dynamical variable. We derive the Lorentz primary constraints, which are present in TEGR and, with a slight modification, in $f(\mathbb T)$ gravity. We show subtleties in the computation of the algebra of these constraints among themselves, which come from spatial derivatives of the tetrad field appearing in the non-homogeneous term. Previous works that compute the result of the Lorentz algebra for the pure-tetrad constraints \cite{Maluf:2013gaa,Mitric:2019rop,Blagojevic:2020dyq} do not show these subtleties with enough detail. We also use some toy models to better understand how to perform computations with spatial derivative terms of the fields.

Later we introduce the covariant formalism of teleparallel gravity. That is, a formulation that allows to drop off the assumption of vanishing spin connection, and introduces a spin connection defined in terms of Lorentz matrices, such that if Lorentz rotated at the same time as the tetrad, the theory remains invariant. We derive the primary constraints associated with such transformation, and prove that they commute among themselves and with the previously found pure-tetrad Lorentz constraints, if covariantised.  These constraints are precisely the ones looked for in the paper \cite{Maluf:2018coz}, where it is argued that the covariant approach would be deficient since the new degrees of freedom introduced by arbitrary Lorentz matrices are not compensated by additional primary constraints. 

A common claim that can be found in the teleparallel community is that the covariant approach solves the problem of violation of local Lorentz invariance in $f(T)$ gravity and similar models \cite{TeleWorSe,Bahamonde:2021gfp}. As we argue here, such claim is wrong, and it is originated from misunderstanding a couple of basic ideas that we support throughout this paper:
\begin{itemize}
    \item ${\bf (i)}$ there are two different types of local Lorentz transformations (LLT), and
    \item ${\bf (ii)}$ the violation of Lorentz invariance occurs in only one of them.
\end{itemize}
In order to support the point ${\bf (i)}$, we have proved that simultaneous LLT are a gauge symmetry of any modified teleparallel gravity. This is because they are associated with primary constraints that appear almost trivially in the formalism. These constraints commute with all the remaining primary constraints present in the theory. As there is no serious reason to think that they would not commute with the Hamiltonian and momenta constraints coming from diffeomorphism invariance (which is always present in modified teleparallel), then simultaneous LLT primary constraints can certainly be regarded as first class.

A different story must be told about Lorentz transformations performed only in the tetrad field. These transformations are not anymore a gauge symmetry, except for very particular cases when there are remnant symmetries of the Lorentz group \cite{Ferraro:2014owa}. The Lorentz transformations performed only on the tetrad is still allowed, even if not a gauge symmetry. In the Hamiltonian picture we see that such transformations are associated with primary constraints that are naturally obtained once the momenta are defined.

The question now is which of the two kinds of Lorentz transformations is broken when passing to nonlinear TEGR models, that is point {\bf (ii)}. The pure-tetrad Lorentz transformations form the  closed algebra of the generators of the Lorentz group only in the case that we are considering TEGR. As it has been shown in some papers \cite{Li:2011rn,Blagojevic:2020dyq}, and as it can be deduced from our own computations here, the algebra does not close for a nonlinear modification of TEGR, and Lorentz symmetries are partially lost in a way not thoroughly understood yet. The introduction of the simultaneous LLT does not fix the algebra by making it closed, therefore it is irrelevant as a tool for making teleparallel theories well defined.

One might doubt whether the pure-tetrad Lorentz symmetry is important. However, even if it is not, the covariantisation does not change anything about the formal troubles of breaking this symmetry and producing the Poisson brackets' algebra of non-constant rank and therefore not well defined number of degrees of freedom. Neither are philosophical issues of ``preferred frame'' effects really solved. The possible non-equivalent teleparallel solutions with the same metric remain available in their covariant versions, too \cite{Golovnev:2021lki}.

The considerations presented in this work can be used for any theory written in the teleparallel framework, whose Lagrangian is built in terms of the torsion tensor as building blocks. The simultaneous LLTs are gauge symmetries of Lorentz-covariant models. A direct proof that the their primary constraints are first class will depend on the Lagrangian under consideration, because for each one of them the Hamiltonian constraint changes, and therefore its Poisson bracket with these constraints too. This formal proof will be presented elsewhere. In the meantime, we present useful mathematical tools to extend the Hamiltonian analysis of covariant teleparallel models, and contribute to a better theoretical understanding of them.

\section*{Acknowledgments}
The authors thank Adel Awad, Sebasti\'an Bahamonde, Milutin Blagojevi\'c, Daniel Blixt, Rafael Ferraro, James M.  Nester, and participants of TeleWorSe 2021 workshop-seminar \cite{TeleWorSe} for numerous discussions on theoretical viability of modified teleparallel gravity  and Hamiltonian formalism in the teleparallel framework. MJG was funded by the Estonian Research Council grant MOBJD622, and by FONDECYT-ANID postdoctoral grant 3190531.

\end{document}